\documentstyle[12pt]{article}

\begin{document}
\title{\bf VARIABLE COEFFICIENT THIRD ORDER KdV TYPE OF EQUATIONS}
\author{ Metin G{\" u}rses\\
{\small Department of Mathematics, Faculty of Sciences}\\
{\small Bilkent University, 06533 Ankara - Turkey}\\
{\small and}\\
Atalay Karasu\\
{\small Department of Physics , Faculty of Arts and  Sciences}\\
{\small Middle East Technical University , 06531 Ankara-Turkey}}
\begin{titlepage}
\maketitle
\begin{abstract}
We show that the integrable subclassess of the equations\\
$q_{,t}=f(x,t)\,q_{,3}+H(x,t,q,q_{,1})$ are the same as the
integrable subclassess of the equations $q_{,t}=q_{,3}+F(q,q_{,1})$.
\end{abstract}
\end{titlepage}

Classification of nonlinear partial differential equations
possessing infinitely many symmetries in 1+1 dimensions  has started
almost two decades ago.
So far the complete classification has been done for some evolution
type of autonomous equations\cite{MIK,IBR,FOK,SIV}. There are some partial
attempts of the classification of the non-autonomous type of equations
\cite{MIK,SIV1,SIV2,OLV,HER}.  In 1+1 or 2+0 dimensions
almost all definitions of integrability coincide. But what is
important is the ease of applicability. Recently we have introduced a new
approach which is based on the compatibility of the symmetry equation
(linearized equation) and an eigenvalue equation \cite{GUR}. Our method
can be put into an algorithmic scheme and utilized for two purposes.
The first
is to test whether a given partial differential equation is integrable.
The second is to classify nonlinear partial differential equations
according to whether they admit generalized symmetries.

In this work we show that the most general equations of the type
$q_{,t}=f(x,t)\,q_{,3}+H(x,t,q,q_{,1})$ , up to coordinate
transformations , have the same integrable subclass as the autonomous equations
$q_{,t}=q_{,3}+F(q,q_{,1})$. Here $f(x,t)$ is an analytic function of the
independent variables $x$ and $t$ , $H$
is a function of the dependent variable $q$ , its $x$-derivative $q_{,1}$
and also on the independent variables $x$ and $t$ . The function $F$
depends on only $q$ and $q_{,1}$. First we will give an outline of
the method

Consider an evolution equation of the form

\begin{equation}
q_{,t}=K(x,t,q,q_{,1},q_{,2},...,q_{,n})\equiv K(q),  \label{a1}
\end{equation}

\noindent
where $q_{,i}=(\frac{\partial}{\partial x})^{i}q$,~~~~~~$i=0,1,2,...,n$.
The order of $ K(=n)$ is called the order of the equation.
A symmetry $\sigma(x,t,q) $  of Eq.(\ref{a1}) satisfies

\begin{equation}
\sigma_{,t}=K^{\prime}(\sigma)=\sum_{i=0}^{n}\frac{\partial K}{\partial
q_{,i}}\sigma_{,i}                             \label{a2}
\end{equation}

\noindent
such that Eq.(\ref{a1}) is form invariant under the transformation

\begin{equation}
q\rightarrow q+\epsilon \sigma,\;\;\;\;\;\;\;\;\;     (\epsilon,
\mbox{infinitesimal})  \label{a3}
\end{equation}

\noindent
Here $\sigma(x,t,q)$ is a differentiable function of $q,q_{,1},q_{,2}$,....
and the prime denotes the Fr\'{e}chet derivative.

In \cite{GUR} we conjectured that a nonlinear partial differential equation
is integrable if the linearized equation(\ref{a2})  supports an eigenvalue
equation. Therefore let us introduce an eigenvalue equation
, linear in $\lambda$, for $\sigma$ in the form

\begin{equation}
\sigma_{,n}=\sum_{i=0}^{n-1}(A_{i}\lambda+B_{i})\sigma_{,i} ,  \label{a4}
\end{equation}

\noindent
where $A_{i} $ and $B_{i}$ are functions of $x$, $t$, and $q_{,i}$. Their
dependences on $q_{,i} $ are  decided by the order of $K$. The order
of the eigenvalue equation is determined by the order of $K$.
The compatibility of linearized and eigenvalue equations, at all powers
of $\lambda$, gives:

{\bf a)} a set of algebraic equations among $A_{i}$, $B_{i}$ and
$\frac{\partial K}{\partial q_{,i}}$'s ;

{\bf b)} a set of coupled PDE's among $A_{i}$, $B_{i}$ and
$\frac{\partial K}{\partial q_{,i}}$'s.

\noindent
Using the definition of total derivatives

\begin{eqnarray}
&&\frac{df}{dx}=D_{x}f=\frac{\partial f}{\partial x}
         +\sum_{i=0}^{\infty} q_{,i+1} \frac{\partial f}{\partial q_{,i}},
\label{a5} \\
&&\frac{df}{dt}=D_{t}f=\frac{\partial f}{\partial t}
        +\sum_{i=0}^{\infty} K_{,i}\frac{\partial f}{\partial q_{,i}},
 \label{a6}
\end{eqnarray}

\noindent
for any function $f$ in the set {\bf b} of coupled PDE's and comparing
coefficients of $q_{,i}$'s , we obtain several classes of $A_{i}$,
$B_{i}$ along with the explicit forms of $K$ in a self consistent way.
If the integrability is proved for a given class, the eigenvalue equation
(\ref{a4}) can always be put in the form

\begin{equation}
M\sigma=\lambda N\sigma,        \label{a7}
\end{equation}

\noindent
where  $M$ and $N$ are local operators and
depend on $x, t, q_{,i}$. Equation (\ref{a7}) is nothing but the definition of
the recursion operator, provided that $N^{-1}$ exists,

\begin{equation}
R=N^{-1}M,              \label{a8}
\end{equation}

\noindent
which maps symmetries to symmetries

\begin{equation}
R\sigma_{n}=\sigma_{n+1},    \label{a9}
\end{equation}

\noindent
where $n$ is a non-negative integer. Thus the existence of an eigenvalue
equation
(\ref{a4}) is equivalent to the existence of a recursion operator.

As an illustration let us  give the classification of third
order autonomous evolution  equations of the form

\begin{equation}
q_{,t}=q_{,3}+F(q,q_{,1}).         \label{a11}
\end{equation}

\noindent
This classification has been investigated by several authors, mainly
from the point of view of their integrability\cite{MIK,IBR1,FOK}.
Let us follow the method outline above.

\noindent
{\bf Linearized equation:}

\begin{equation}
\sigma_{,t}=\sigma_{,3}+\frac{\partial F}{\partial q_{,1}}\sigma_{,1}+
\frac{\partial F}{\partial q}\sigma.       \label{a12}
\end{equation}

\noindent
{\bf Eigenvalue equation:}

\begin{equation}
\sigma_{,3}=(A_{2}\lambda+B_{2})\sigma_{,2}+(A_{1}\lambda+B_{1})\sigma_{,1}
+(A_{0}\lambda+B_{0})\sigma,              \label{a13}
\end{equation}

\noindent
where $A_{i}$ and $B_{i}$ depend on $q$, $q_{,1}$ and $q_{,2}$.
The compatibility equation of Eq.(\ref{a12}) and Eq.(\ref{a13}) gives
the following integrable equations with non zero eigenvalue coefficients
\begin{equation}
\mbox{{\bf Case I}.}~~~~~~q_{,t}=q_{,3}+ {a\,\over 6}q_{,1}^{3} +
{b \,\over 2}q_{,1}^{2}+c\,q_{,1}+d,        \label{a14}
\end{equation}

\noindent
with

\begin{eqnarray}
&&B_{2}=\frac{a q_{,2}}{a q_{,1}+b} \;,~~~~~
B_{1}=-\frac{1}{3}\left[a q_{,1}^{2}+2 b q_{,1}+2 c\right],
A_{1}=1 \nonumber\\
&& B_{0}=\frac{1}{3(aq_{,1}+b)} \left[q_{,2}(2ac-b^2)\right]\;,~~~~~
 A_{0}=\frac{a q_{,2}}{a q_{,1}+b} .    \label{a15}
\end{eqnarray}

\noindent
Here $a$,$b$ and $c$ are constants, and

\begin{equation}
\mbox{\bf Case II.}~~~~q_{,t}=q_{,3}+{a \, \over 6}q_{,1}^{3}+b(q)
q_{,1},\;~~~~~
3\frac{d^3 b}{d q^3}+4 a \frac{d b}{d q} =0,   \label{a16}
\end{equation}

\noindent
with

\begin{eqnarray}
&&B_{0}=\frac{1}{3} \left[2\frac{b q_{,2}}{q_{,1}}-3 \frac{db}{dq} q_{,1}
\right], \;~~~~~~B_{1}=-\frac{1}{3}(aq_{,1}^{2}+2 b),\;~~~B_{2}=\frac{q_{,2}}
{q_{,1}},  \nonumber\\
&& A_{0}=-B_{2},\;,~~~A_{1}=1.\;      \label{a17}
\end{eqnarray}

\noindent
Here $a$ is a constant. The basic equations in the classification are the
KdV ($a=0$, $b=6q$, in case II), pKdV ($a=0$, in case I), mKdV
($a=0$, $b=6q^{2}$, in case II),
pmKdV($b=0$, in both cases) and CDF equation ($a=-{3\, \over 4}$, in case II).
The recursion operators for equations (\ref{a14}) and (\ref{a16}) are found
by the utility  (\ref{a15}) and (\ref{a17}). They are respectively given by

\begin{eqnarray}
\mbox{\bf I.}~~~&&R=D^{2}+{2c \over 3}+a\,{q_{,1}^{2}\,\over
3}+{2bq_{,1}\,\over 3}
-{aq_{,1}\,\over 3}D^{-1}(q_{,2} .)-{b\,\over 3}D^{-1}(q_{,2} .) \label{a18}\\
\mbox{\bf II.}~~~&&R=D^{2}+{aq_{,1}^{2}\,\over 3}+{2b\,\over 3}-{a\,q_{,1}
\over 3}D^{-1}(q_{,2}.)
+{q_{,1} \over 3}D^{-1}(\frac{db}{dq} .),    \label{a19}
\end{eqnarray}

\noindent
where $D^{-1}=\int_{-\infty}^{x} .dx$. We have the following proposition.

{\bf Proposition 1}:\cite{MIK,IBR,FOK} Under the symmetry classification
the equations of the type $q_{,t}=q_{,3}+ F(q,q_{,1})$ ,
up to coordinate transformations ,
has  integrable subclassess given in (\ref{a14}) and (\ref{a16}).

Our aim ,in this work, is to give a classification of
 the non-autonomous type of integrable equations (\ref{a11})

\begin{equation}
q_{,t}=f(x,t)q_{,3}+H(x,t,q,q_{,1})  \label{a24}
\end{equation}

\noindent
We divide the classification procedure, for Eq.(\ref{a24}),
into the following three  cases:
{\bf i)} $f$ depends only on $t$ , {\bf ii)} $f$ depends  only on $x$ , and
{\bf iii)} $f$ depends on both $x$ and $t$.

\noindent
{\bf i})$f$ depends only on $t$ :
One of the integrable subclass ,using our classification scheme, is

\begin{equation}
q_{,t}=v^{3}\,q_{,3}+({h\,w^{2}\,q^{2} \over 2}\,+{c_{1}\,w\,h
\,v\,}q+hc_{2}\,v)
 \,q_{,1}+({\dot{h}\,\over 2\,h}-{\dot{v}\,\over v})
 x\,q_{,1} -{\dot{w}\,q\,\over w},\label{a26}
\end{equation}

\noindent
where $f(t)=v(t)^{3}$ , $h$, $w$ depend on $t$ only and
$c_{1}$, $c_{2}$ are constants. The dot appearing
over a quantity denotes  $t$ derivative. The recursion operator is given by

\begin{equation}
R={v^{2}\,\over h}\,D^{2}+{w^{2}\,\over 3}\,q^{2}+{2\,c_{1}\,\over 3}\,q
+{2\,\over 3}\,c_{2}+{w^{2}\,\over 3}q_{,1}\,D^{-1}(q\,.)\,+{c_{1}\,w\,\over
3}\,
q_{,1} D^{-1}. \label{a27}
\end{equation}

\noindent
We observe that Eq.(\ref{a26}) is transformed into an equation which belongs
to the case II ($a=0$) in Eq.(\ref{a16}) through the transformation

\begin{eqnarray}
&&q=w^{-1}u(\xi,\tau) ,  \nonumber \\
&&\xi=x \beta(t),\,~~~~\beta=\frac{h^{1/2}}{v}
,~~~~\tau=\int^{t}\,{ h^{3/2}\,dt^{\prime}}.  \label{a30}
\end{eqnarray}

\noindent
In the classification programs, if it is possible, we
transform (by coordinate or contact transformations)
the given class of PDE's to more simpler ones.
To this end in the sequel we shall
transform all cases ({\bf i}, {\bf ii}, {\bf iii}) to the form

\begin{equation}
q_{,\tau}=q_{,3}+H_{2}(\xi,\tau,q,q_{,1})  \label{a50}
\end{equation}

\noindent
and classify this type of equations. In this first case ({\bf i}) we have the
following proposition.

\noindent
{\bf Proposition 2}: Under the symmetry classification the equations
of the type

\begin{equation}
q_{,t}=f(t)\,q_{,3}+ F_{1}(x,t,q,q_{,1}) , \label{a31}
\end{equation}

\noindent
up to coordinate transformations ,
give the same integrable subclass as in propopsition 1.
Furthermore, Eq.(\ref{a31}) reduces to Eq.(\ref{a50})
by the transformation $dt=\frac{1}{f}d\tau$ and $x=\xi$.

\noindent
{\bf ii}) $f$ depends only on $x$ : In this case the form of equation is

\begin{equation}
q_{,t}=f(x)q_{,3}+F_{2}(x,t,q,q_{,1}),   \label{a33}
\end{equation}

\noindent
one integrable class turns out to be simple

\begin{equation}
q_{,t}= q_{,3}+q_{,1}^{2}+c_{1}x +c_{2} , \label{a34}
\end{equation}

\noindent
Recursion operator for Eq.(\ref{a34}) is

\begin{equation}
R=D^{2}+\frac{4}{3}q_{,1}-\frac{4}{3}c_{1}t-\frac{2}{3}D^{-1}(q_{,2} .),
\label{a35}
\end{equation}

\noindent
Now, differentiate Eq.(\ref{a34}) with respect to $x$ and substitute
$q=z_{,1}$ and use the transformations $x=\xi-\frac{1}{2}c_{1}t^{2}$
, $t=\tau$ then Eq.(\ref{a34}) belongs to Eq.(\ref{a14}). Before
proceeding to the next case, we observe the following:

\noindent
{\bf Proposition 3}: Under the symmetry classification
the equations of the type (\ref{a33}) up to coordinate transformation ,

\begin{eqnarray}
q&=&f^{1/3}u(\xi,\tau),
 \nonumber\\
\xi&=&\int^{x}\frac{1}{f^{1/3}}dx^{\prime},\,~~~~~~~~~~~\tau=t, \label{a52}
\end{eqnarray}

\noindent
give the same integrable subclass as in Eq.(\ref{a50}).

{\bf iii}) $f$ depends on both $x$ and $t$: In this case we have
 $q_{,t}=f(x,t)\,q_{,3}+F_{3}(x,t,q,q_{,1})$ type of equations
and its one integrable class turns out to be
relatively simple. Below we give this equation and its recursion operator

\begin{equation}
q_{,t}=u^{3}\,q_{,3}+\left [{a\, \over 2\,u}q^{2}\,+{3 \over 2}\,u\,(u_{,1}^{2}
-2\,u\,u_{,2}) \right ]\,(q_{,1} -{u_{,1} \over u}\,q) \label{a37}
\end{equation}

\noindent
where $a$ is an arbitrary constant . Here we have set $f(x,t)=u(x,t)^{3}$ and
$u(x,t)$ satisfies the Harry Dym equation

\begin{equation}
u_{,t}=u^{3}\,u_{,3}     \label{a38}
\end{equation}

\noindent
It means that Eq.(\ref{a37}) is integrable if $u$ satisfies the Eq.(\ref{a38}).
The recursion operator is given by

\begin{equation}
R=u\,s\,D^{-1}\,u\,D\,({1 \over s}\,R_{1}.\,) ,  \label{a39}
\end{equation}

\noindent
where

\begin{eqnarray}
R_{1}&=&D^{2}-{u_{,1}\,q_{,2} \over us}+{u_{,2}\,q_{,1} \over us}-{u_{,2} \over
u}
+{a\,q^{2} \over 3\,u^{4}}+s\,D^{-1}\,({V \over s}.),  \nonumber  \\
V&=&{a \over 3\,u^{4}}\,q\,q_{,1}+{1 \over us} \left [u_{,1}\,q_{,3}-
q_{,1}\,u_{,3} +{2\,s_{,1} \over s}\,(-u_{,1}\,q_{,3}+q_{,1}\,u_{,3}) \right ]
\label{a40}
\end{eqnarray}

\noindent
and $s=-q_{,1}+q\,{u_{,1} \over u}$.
Eq.(\ref{a37}) together with (\ref{a38}) is equivalent to the mKdV.
We will give the proof of this in two steps:
Let $f=u^{3}$ and $q=u\,z$ then we have

\begin{equation}
z_{,t}=u^{3}\,z_{,3}+3\,u^{2}\,u_{,1}\,z_{,2}+ ({a\,\over 2}u z^{2}+{3\,\over
2}uu_{,1}^{2})z_{,1},  \label{a41}
\end{equation}

\noindent
where $z(x,t)$ is the new dynamical variable.
Now let us perform the following transformation

\begin{equation}
\xi=\int^{x}\, {d\,x^{\prime} \over u{(x^{\prime},t})}~~,~~ \tau=t \label{a42}
\end{equation}

\noindent
It is straightforward to show that under this transformation Eq.(\ref{a41})
goes to the mKdV. Now we state the following proposition.

{\bf Proposition 4}: Under the symmetry classification the equations of the
type
$ q_{,t}=f(x,t)\,q_{,3}+F_{3}(x,t,q,q_{,1}) $ , up to the
transformations ,

\begin{eqnarray}
q&=&f^{1/3}z(\xi,\tau), 
\nonumber \\
\xi&=&\int^{x}\frac{1}{f^{1/3}}dx^{\prime},\,~~~~~~~~~~~\tau=t, \label{a54}
\end{eqnarray}

\noindent
like in the previous example, give the  same integrable
subclass ( for $z(\xi,\tau)$) as in Eq(\ref{a50}).

Hence whatever the coefficient function $f(x,t)$ we showed that in general
the type of equations (\ref{a24}), by  coordinate
transformations reduce to the following type of equations

\begin{equation}
q_{,t}=q_{,3}+P(x,t,q,q_{,1})    \label{a55}
\end{equation}

\noindent
We now give the classification of this type of equations.

\begin{eqnarray}
\mbox{\bf (1)}~~~  &&q_{,t}=q_{,3}+{a\,\over 2}q_{,1}^{2}+b
q_{,1}-{\dot{w}\,\over w}q
+c ~~~~~~~~~~~~~~~~~~~~~~~~~~~~~~~~~~~\label{a56}\\
\noindent
\mbox{with} \nonumber\\
&&\dot{b}=b_{,3}+{b b_{,1}}-a c_{,1}+{\dot{h}\,\over 2h}b
+{\dot{d}\,\over 2 h}-{\dot{h}\, \over h^{2}}d+{({\ddot{h}\,\over 2h}-
{\dot{h}^{2}\,\over h^{2}})x}  \label{a57}
\end{eqnarray}

\noindent
where $w={a\,\over\sqrt{h}}$ ,$a$,$d$,$h$ depend on $t$ only and $b$, $c$
depend on $x$, $t$.

\begin{equation}
\mbox{\bf (2)}~~~ q_{,t}=q_{,3}-{a^{2}\,\over 8}q_{,1}^{3}+{({h\,\over w}e^{aq}
-whe^{-aq}+b)}q_{,1}
+{\dot{h}\,\over 2h}xq_{,1}-{\dot{a}\, \over a}q+{\dot{w}\,\over aw} \label{
a58}
\end{equation}

\noindent
where $a$, $w$, and $b$ depend on $t$ only.

\begin{equation}
\mbox{\bf (3)}~~~q_{,t}=q_{,3}+{a\,\over 2}q^{2}q_{,1}
+({b\,\over w})^{1/2}qq_{,1}+
{({\dot{h}\,\over 2h}x+c)}q_{,1}+{\dot{w}\,\over 2w}q-
{h\,\over 2}({w\,\over b})^{1/2}\dot{d}  \label{a59}
\end{equation}

\noindent
where $b=h^{2}d$, and all parameters appearing  in the equation depend on
$t$.

\begin{eqnarray}
\mbox{\bf (4)}~~~ q_{,t}&=&q_{,3}+{a\, \over 2}q^{2}q_{,1}+bqq_{,1}
-{1\,\over 2}(-{\dot{h}\,\over h}x
-{b^{2}\,\over a}+{c\,\over ah})q_{,1} +
{b_{,1}\,\over 2}q^{2} ~~~~~~~~~~~~\nonumber \\
&-&{1\,\over 2}({\dot{w}\,\over w}-{2b\,\over
a}b_{,1})q
-{1\,\over 2 a^{2}}({-\dot{a}b}-{2ab_{,3}}-{\dot{h}\,
\over h}b_{,1}ax \nonumber\\
&-&{b^{2}}b_{,1}+{c\,\over h}b_{,1}
+{2\dot{b}a}-
{\dot{h}\,\over h}ab)   \label{a60}
\end{eqnarray}

\noindent
where $w={a\,\over h}$, $a$, $h$,$c$ depend on $t$ only and
$b$ depends on $x$, $t$.

\begin{eqnarray}
\mbox{\bf (5)}~~&&~~ q_{,t}=q_{,3}+aqq_{,1}
+bq_{,1}-{1\,\over 2}({\dot{w}\,\over w}-2b_{,1})q ~~~~~~~~~~~~~~~~~~~~~~~~~~~
{}~~~~~~~~~ \nonumber\\
&&-{1\,\over 2 a}(-2b_{,3}-2bb_{,1}+2\dot{b}-{\dot{c}\,\over h}-{\ddot{h}\,
\over h}x+2{\dot{h}^2\,\over h^2}x-{\dot{h}\, \over h}b+2{\dot{h}\, \over h^2
}c) \label{a61}
\end{eqnarray}

\noindent
where $w={a^{2}\,\over h}$, $a$ depend on $t$ only and $b$ depends on
$x$, $t$.

\begin{eqnarray}
\mbox{\bf (6)}~~~~ q_{,t}=q_{,3}+{a\,\over 6}q_{,1}^{3}+{ab\,\over 2}q_{,1}^{2}
+{1\,\over 2}
({\dot{h}\, \over h}x
+b^{2}a-{c\,\over h})q_{,1}
-{1\,\over 2}({\dot{a}\,\over a}q-2d)~~~~~  \label{a62}
\end{eqnarray}

\noindent
with
\begin{equation}
d_{,1}={1\,\over 2}[ 2b_{,3}+b_{,1}({\dot{h}\,\over h}x+ab^{2}
-{c\,\over h})
-2\dot{b}-{b\dot{a}\,\over a} +{b\dot{h}\,\over h}]    \label{a63}
\end{equation}

\noindent
where $b$, $d$ depend on $t$, $x$ and $a$, $c$ and $h$ depend on $t$ only.

\begin{eqnarray}
\mbox{\bf (7)}~~~~q_{,t}&=& q_{,3}-{a^2\,\over 8}q_{,1}^{3}
-{3 a b_{,1}\, \over 8 b}q_{,1}^{2}
+({b \over a}e^{aq}-{ah^2f\,\over b}e^{-aq}
+{\dot{h} \, \over 2 h}x~~~~~~~~~~\nonumber \\
&+&{c\,\over 2 h}-{h d\,\over 2} -{3b_{,1}^2\,\over 8 b^2})q_{,1}
-{\dot{a}\,\over a}q
+({b_{,1}\,\over a^2}e^{aq}-{h^2 fb_{,1}\,\over b^2}e^{-aq}\nonumber \\
&+&{\dot{h}b_{,1}\over 2abh}x+{b_{,3}\,\over ab}-{3b_{,1}b_{,2}\,\over ab^2}
+{15b_{,1}^3\,\over 8 a b^3}+{cb_{,1}\,\over 2abh}-{dhb_{,1}\,\over 2ab}
+{\dot{w}\,\over aw})    \label{a64}
\end{eqnarray}

\noindent
where  $w={ah\over b}$, $a$,$c$ depend on $t$ only, $b$ depends on $x$
,$t$ and $d$, $f$ are constants.
All these classess are transformable to those given in (\ref{a14})
and (\ref{a16}).For this purpose , we first perform the following
transformation

\begin{equation}
q=\alpha(t) z(x,t)+\beta(x,t),     \label{a65}
\end{equation}

\noindent
where $\alpha$ and $\beta$ are arbitrary functions, and $z(x,t)$ is the new
dynamical
variable. By choosing $\alpha$ and $\beta$ properly
we eliminate arbitrary functions of $x$ and $t$
appearing in these equations. Secondly
,if the resultant equations contain further arbitrary functions
depending upon $t$, we perform
the transformation of the following type

\begin{equation}
z=v(x,t)+s_{0}x^{2}+s_{1}x+s_{2},     \label{a66}
\end{equation}

\noindent
to eliminate such arbitrary functions as the products of $x^{2}$ and,
$x$ in these
equations. Here $s_{0}$,$s_{1}$ and $s_{2}$  depend only on $t$
and $v(x,t)$ is now the new dynamical variable.
At this point we transform dependent and independent variables according to

\begin{eqnarray}
v(x,t)&=&\mu(t) u(\xi,\tau), \nonumber\\
\xi&=&x\rho(t)+\gamma(t) ,~~~~~~~~~~~~~~\tau=\nu(t)  \label{a67}
\end{eqnarray}

\noindent
which reduces the classess(1-7) to one of the type given in equations
 (\ref{a14}) and (\ref{a16}) exactly.
As an example Eq.(\ref{a60}) is transformed into the Eq.(\ref{a14})
through the following transformations

\begin{eqnarray}
&&q={h^{1/2}\,\over a}u(\xi,\tau)-{b \over a},\nonumber \\
&&\xi=xh^{1/2}+\gamma(t), ~~\tau=\int^{t} h^{3/2}dt^{\prime},
\end{eqnarray}

\noindent
where

\begin{equation}
\gamma=-\int^{t}({c\,\over 2ah^{1/2}}+h^{2/3})dt^{\prime}.   \label{a68}
\end{equation}

\noindent
Finally we have the following proposition.

\noindent
{\bf Proposition 5:} Under the symmetry classification the
integrable subclass of the type of equations (\ref{a55})
is, up to coordinate  transformations, equivalent to the
equations (\ref{a14}) and (\ref{a16}).

In conclusion this work shows that there are no generic integrable
non-autonomous type of equations (\ref{a24}). Any integrable PDE
(admiting infinitely many generalized symmetries)
containing explicit ($x ,t$) dependencies of the form (\ref{a24})
is transformable into (\ref{a11}).

\noindent
This work is partially supported by the Scientific and Technical
Research Council of Turkey (TUBITAK).

\end{document}